\documentclass[conference]{IEEEtran}
\IEEEoverridecommandlockouts
\usepackage{amsmath,amssymb,amsfonts}
\usepackage{algorithmic}
\usepackage{graphicx}
\usepackage{textcomp}
\usepackage{xcolor}
\usepackage[style=ieee, maxnames=6, minnames=1, doi=false, isbn=false, url=false]{biblatex}
\usepackage{threeparttable}
\usepackage{multirow, multicol}
\usepackage{threeparttable}
\usepackage{booktabs}
\usepackage[a4paper, total={184mm,239mm}]{geometry}
\def\BibTeX{{\rm B\kern-.05em{\sc i\kern-.025em b}\kern-.08em
    T\kern-.1667em\lower.7ex\hbox{E}\kern-.125emX}}
\begin{document}

\title{VRank: Enhancing Verilog Code Generation from Large Language Models via Self-Consistency\\
}

\author{
\IEEEauthorblockN{Zhuorui Zhao\textsuperscript{1}, Ruidi Qiu\textsuperscript{1}, Ing-Chao Lin{2}, Grace Li Zhang\textsuperscript{3}, Bing Li\textsuperscript{4}, Ulf Schlichtmann\textsuperscript{1}}
\IEEEauthorblockA{\textsuperscript{1}\textit{Chair of Electronic Design Automation, Technical University of Munich (TUM)}, Munich, Germany \\
\textsuperscript{2}\textit{Computer Architecture and IC Design, National Cheng Kung University} \\
\textsuperscript{3}\textit{Hardware for Artificial Intelligence Group, Technical University of Darmstadt} \\
\textsuperscript{4}\textit{Research Group of Digital Integrated Systems, University of Siegen} \\
Email: $\{$zhuorui.zhao, r.qiu, ulf.schlichtmann$\}$@tum.de, iclin@csie.ncku.edu.tw, grace.zhang@tu-darmstadt.de, bing.li@uni-siegen.de}
}

\maketitle

\begin{abstract}
Large Language Models (LLMs) have demonstrated promising capabilities in generating Verilog code from module specifications. To improve the quality of such generated Verilog codes, previous methods require either time-consuming manual inspection or generation of multiple Verilog codes, from which the one with the highest quality is selected with manually designed testbenches. 
To enhance the generation efficiency while maintaining the quality of the generated codes, 
we propose \textit{VRank}, an automatic framework that generates Verilog codes with LLMs. In our framework, 
multiple code candidates are generated with LLMs by leveraging their probabilistic nature. Afterwards, we group
Verilog code candidates into clusters based on identical outputs when tested against the same testbench, which is also generated by LLMs. Clusters are ranked based on the consistency they show on testbench. To determine the best candidate, Chain-of-Thought is further applied to select the best candidate from the top-ranked clusters. 
By systematically analyzing diverse outputs of generated codes, VRank reduces errors and enhances the overall quality of the generated Verilog code. Experimental results on the VerilogEval-Human benchmark demonstrate a significant 10.5\% average increase in functional correctness (pass@1) across multiple LLMs, demonstrating VRank’s effectiveness in improving the accuracy of automated hardware description language generation for complex design tasks.
\end{abstract}


\section{Introduction}
\label{sec:Introduction}
As chip design complexity escalates alongside the demand for enhanced computational performance, hardware engineers are increasingly confronted with intricate design challenges. Large Language Models (LLMs) have made notable breakthroughs in software code generation~\cite{li_competition-level_2022,xu_automated_2024} and are gradually being applied to hardware code generation \cite{chen_machine_2023,xu_llm-aided_2024}.  For example, \cite{chang_chipgpt_2023,qiu_correctbench_2024} employs LLM to generate Verilog modules from module specifications and test benches, while \cite{blocklove_chip-chat_2023} utilizes LLM as copilot to assist in writing Verilog code for CPU. These efforts highlight LLMs' potential in translating natural language specifications into hardware description languages such as Verilog. 

Despite the promising automation of Verilog generations, the quality of the Verilog codes generated by LLMs is still low due to the limited amount of hardware description data. The issue persists even as researchers gather more Verilog data to train models like~\cite{pei_betterv_2024}. To further enhance the quality of such codes, previous work retried generation for several times till a good enough Verilog is recognized by human engineer~\cite{blocklove_chip-chat_2023}, or used iterative feedback given by human-designed testbench~\cite{ho_verilogcoder_2024}.
Although LLMs are capable of generating correct Verilog codes for specific tasks within several retries, previous work is unable to differentiate which Verilog code sample is better without human effort.

To address the limitations in the previous work, we introduce \textbf{VRank}, an automatic framework designed to improve the efficiency and accuracy of Verilog code generation using LLMs. VRank leverages the probabilistic nature of LLMs to generate multiple code candidates and automatically clusters them based on their functional outputs when tested against LLM-generated testbenches. By ranking these clusters according to their consistency and applying a chain-of-thought reasoning process, VRank selects the code candidate with the highest quality from the top-ranked clusters. Our approach not only enhances the quality of the generated Verilog code but also enables an automatic generation flow without the need for any human intervention. 

The contribution of our paper is summarized as follows:

\begin{itemize}
    \item We propose an automated Verilog generation framework that enhances both the quality of the generated code and the efficiency of the generation process, eliminating the need for human intervention. 
    \item The proposed framework takes advantage of the probabilistic characteristics of LLMs by generating multiple Verilog codes from the same module specification. Such codes are clustered based on their functionalities for subsequent code selection. 
    \item Inspired by self-consistency in machine translation, the clusters are ranked based on the consistency of their simulation outputs. This clustering and ranking strategy can effectively identify the top candidates that are most likely functionally correct. 
   \item We further deploy chain-of-thought to enhance the selection of codes. By reasoning and summarizing from LLMs, this process automatically differentiates between the top two clusters, improving the accuracy of the final selection.
   \item Experimental results demonstrate the effectiveness of VRank, showing a significant improvement in pass@1 rates across different LLMs. The improvement is stable with a sample size from 5 to 50. These findings highlight the potential of our approach to advance automated Verilog code generation, offering a fully automatic perspective for both design and verification in hardware engineering.
\end{itemize}

The rest of the paper is organized as follows. Section \ref{sec:BackgroundandMotivation} presents an overview of existing work on Verilog generation and outlines the basic intuition behind our approach. Section \ref{sec:Proposed Framework} explains our VRank in detail. The experimental setup is explained in Section \ref{sec:ExperimentalSetup}. Section \ref{sec:ExperimentalResults} provides the research question and experimental results. Section \ref{sec:conclusion} concludes the paper.

\section{Background and Motivation}
\label{sec:BackgroundandMotivation}
\subsection{Large Language Models in Code Generation}

Transformer-based Large Language Models (LLMs) have been widely applied in code generation, starting with the advent of Codex~\cite{chen_evaluating_2021} and AlphaCode~\cite{li_competition-level_2022}. They were initially applied to software code generation, which has been the primary focus of early research in this domain. In evaluating these code generation models, the pass@k metric was introduced. The metric evaluates the rate of any of the first $k$ code candidates are functionally correct and pass the testbench.

Recently, LLMs have also been explored to generate hardware description languages such as Verilog~\cite{he_large_2024}. 
For example, DAVE~\cite{pearce_dave_2020} introduced the first workflow tailored for this purpose. Subsequent works including BetterV~\cite{pei_betterv_2024}, ChipNeMo~\cite{liu_chipnemo_2024}, and others~\cite{thakur_benchmarking_2023}\cite{zhao_codev_2024}\cite{liu_rtlcoder_2024}\cite{cui_origenenhancing_2024}\cite{gao_autovcoder_2024}, have explored LLMs specifically fine-tuned for generating Verilog codes. Such work primarily focused on creating and augmenting Verilog explanations to code datasets. The latest model GPT-4~\cite{openai_gpt-4_2024} has shown the capability of generating Verilog code coherently and the potential of being integrated into code agent in \cite{blocklove_chip-chat_2023}\cite{chang_improving_nodate}. Despite the growing power of these domain-specific or general-purpose LLMs, they all show a noticeable gap between pass@1 and pass@10. This gap indicates that users may either settle for suboptimal results or repeatedly retry generation to obtain an acceptable solution without clear stopping criteria. 

Recent research has sought to improve the quality of Verilog codes generated by LLMs via feedback mechanisms. RTLFixer~\cite{tsai_rtlfixer_2024} introduced an autonomous agent framework that integrates simulator feedback with Retrieval Augmented Generation (RAG)~\cite{lewis_retrieval-augmented_2020} to address syntax errors. It cannot address the functionality quality of Verilog codes. Regarding functional error correction, AutoChip~\cite{thakur_autochip_2024} and VerilogCoder~\cite{ho_verilogcoder_2024} explored debugging of Verilog code using detailed, human-written testbenches. Despite improving functional success rates, these solutions are impractical for real-world applications due to the labor-intensive nature of writing exhaustive testbenches.

Although LLMs can also be used to automatically generate testbenches \cite{qiu_autobench_2024}, the quality of these generated testbenches lags behind that of the generated Verilog modules, indicating wrong reference signals inside such testbenches. Therefore, directly using them to facilitate Verilog generation can lead to low code quality.

\begin{figure}[tb]
    \centering
    \includegraphics[width=1\linewidth]{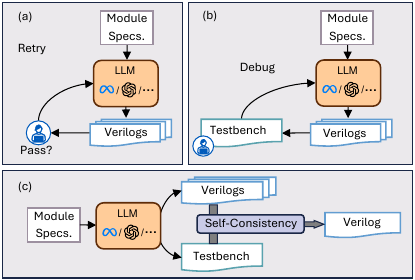}
    \caption{Comparison between (a) direct sampling, (b) debugging, and (c) VRank(proposed). Our methods involve no human-in-loop, while achieving significant improvement on Pass@1 accuracy.}
    \label{fig:comparison}
    \vspace{-0.3cm}
\end{figure}

\subsection{Motivation}

Previous methods for enhancing Verilog code quality either need humans to check if a generated Verilog code passes, as shown in Fig. \ref{fig:comparison}(a), and retry failed generation, or need a testbench written by a human to provide debugging information for further revision, as depicted in  Fig. \ref{fig:comparison}(b). 
Such techniques suffer from either generation inefficiency or time-consuming testbench preparation. To address this challenge, we propose to develop an automatic method that can enhance the quality of generated Verilog, by sampling a number of Verilog code candidates and identifying the highest quality code via self-consistency checks. The concept is shown in Fig. \ref{fig:comparison}(c). 

The basic idea of self-consistency is to select the output sample generated by a model that has the most consistency with the remaining samples. For example, an LLM is used to generate a Verilog module, and it produces five slightly different versions of codes. If three of those versions behave similarly during simulation, one of the three codes can be used as the final version. The rationale behind this selection is that the consistency in code simulation results of the three versions of codes indicates they have a higher probability of being correct compared with the other versions.

Self-consistency can improve the reasoning capabilities of LLMs through majority voting mechanisms, as demonstrated in \cite{wang_self-consistency_2022}. 
It has also been widely used in machine translation~\cite{goel_segmental_2000} and speech-to-text tasks~\cite{fiscus_post-processing_1997}. For instance, in machine translation, Minimum Bayes Risk (MBR) decoding~\cite{goel_segmental_2000}\cite{kumar_minimum_2004} is used to minimize the risk associated with producing a certain translation, where risk is defined as the expected loss between the candidate translation and a set of reference translations. Formally, the MBR decision rule is:
\begin{equation}
\label{equ:MBR}
\hat{y} = \arg\min_{y \in \mathcal{Y}_h} \sum_{y' \in \mathcal{Y}_e} \ell(y, y'),
\end{equation}
where \(y\) is a candidate translation, and \(y'\) is a possible reference translation, which also comes from the model's output. The loss \(\ell(y, y')\), in a form such as BLEU~\cite{papineni_bleu_2002}, punishes candidates that differ greatly from the reference signal. 
Similarly, the Recognizer Output Voting Error Reduction (ROVER)~\cite{fiscus_post-processing_1997} mitigates these issues by selecting outputs that maximize consistency across multiple output samples. 

\begin{figure*}[tb]
    \centering
    \includegraphics[width=1\linewidth]{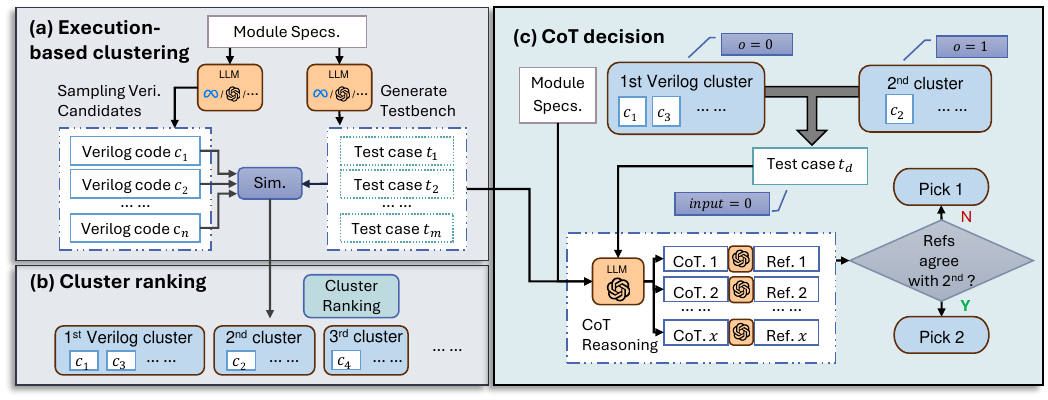}
    \caption{The outline of VRank. Our framework contains three major steps. (a) Execution-based clustering, (b) Cluster ranking, and (c) CoT decision.}
    \label{fig:overall_framework}

    \vspace{-0.6cm}
\end{figure*}

Inspired by these approaches, we apply self-consistency principles to Verilog code generation using two specific evaluation criteria. First, we evaluate the self-consistency of LLM-generated Verilog code candidates based on their outputs when run in an LLM-generated testbench. Second, we assess the self-consistency of LLM-generated reasoning to establish a more reliable reference signal. The details of evaluating these consistencies will be presented in the subsequent section.

\section{Proposed Framework}
\label{sec:Proposed Framework}
The VRank framework is designed to enhance the quality of Verilog code generation through a three-stage process: execution-based clustering, cluster ranking, and Chain-of-Thought final decision, as depicted in Fig. \ref{fig:overall_framework}. \textbf{Execution-based clustering}: we use LLMs to generate multiple code candidates and a testbench that contains multiple test cases. We run a simulation of each code candidate on this testbench. Clusters are formed based on the simulation outputs. \textbf{Cluster ranking}: Clusters are ranked using a scoring function that quantifies consistency among code candidates. \textbf{Chain-of-Thought (CoT) final decision}: Inconsistency between the top clusters is identified, and Chain-of-Thought prompting is employed to finalize the choice of the best candidate. The details of each step are outlined below.

\subsection{Execution-based clustering}
\begin{figure}
    \centering
    \includegraphics[width=0.95\linewidth]{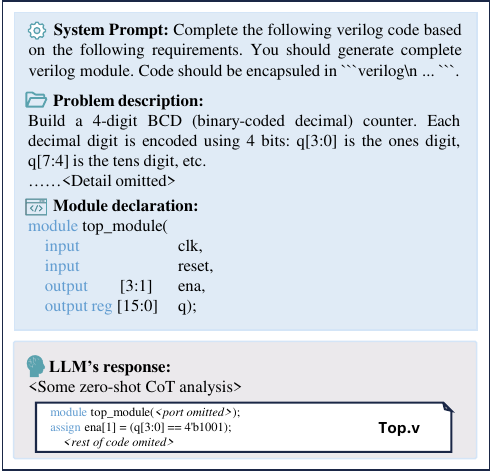}
    \caption{generating Verilog code candidates from LLM}
    \label{fig:prompt}
    \vspace{-0.6cm}
\end{figure}

The VRank process starts by querying a pre-trained large language model (LLM) to generate a set of candidate Verilog codes, denoted as \(C=(c_1, c_2, ...,c_n)\), and a set of test case inputs, \(T=(t_1, t_2, ..., t_m)\). The method used to prompt the LLM for generating candidate codes is shown in Fig. \ref{fig:prompt}. The testbench that contains multiple test cases is obtained using a testbench generation technique similar to that in ~\cite{qiu_autobench_2024}. We query the LLM to produce multiple test cases, then concatenate them into the same testbench $T$. This testbench is designed to only print the input and output on each test case, without providing any reference signal to check the correctness of the value of each output.

We will then run all code candidates on the same testbench and collect the simulation results of each code \((c_1(T), (c_2(T),...)\). By analyzing the outputs of the code candidates across the test cases, we group the candidates that produce identical simulation results on all test cases, indicating that code candidates are clustered based on their functional equivalence.

\subsection{Cluster Ranking and Candidate Selection}
Once the clusters are formed, the next step is to rank the clusters and select the top-ranked clusters to identify the code candidate with the highest quality. The ranking is based on the simulation consistency of code candidates in each cluster. To quantitatively evaluate the consistency, we assign each cluster with a score.

The score evaluation of each cluster is the same with the inconsistency minimization in Minimum Bayes Risk (MBR) decoding. Specifically:
\begin{equation}
\label{equ:MBR}
R(c) = n -\sum_{c' \in \mathcal{C}}\ell_{strict}(c, c'),
\end{equation}
with the loss function
\begin{equation}
    \ell_{strict}(c, c')=\max_{t\in T} \mathbf{1}[c(t)\neq c'(t)]
\end{equation}
where the score of each code candidate $c$ is given by $R(c)$. $R(c)$ represents the \textit{inconsistency} of code candidate $c$ with all other candidates. If $c$ and $c'$ have any inconsistency in the simulation output, it will result in a loss of 1. A candidate with a failed simulation is also defined as inconsistent with any other candidate.
For example, in Fig. \ref{fig:overall_framework}(b), if $n=10$ Verilog candidates are generated by an LLM, cluster 1 has 5 candidates, cluster 2 has 3 candidates and cluster 3 has 2 candidates, this would result in a score of $10-5=5$ for all candidates in cluster 1 because it has difference with 5 other samples; cluster 2 and 3 will have scores of 3 and 2.

After assigning each cluster with a score, we rank clusters according to their scores in descending order. We can select top candidates from top-ranked clusters. For example, we randomly select only one code candidate from each cluster in order, as long as there is at least one candidate in this cluster. 
In this way, we not only ensure that the optimal candidate under MBR decoding is selected, but also avoid sampling functionally equivalent code samples.
We will further apply chain-of-thought reasoning to determine the final candidate, as detailed in Sec. \ref{subsec:CoT}.

\subsection{Chain-of-Thought Reasoning on Inconsistency}
\label{subsec:CoT}
After clustering and ranking, we obtain an ordered list of code candidate clusters. Top-ranked clusters always show output inconsistencies on certain test cases, making it possible to identify the correct candidate by analyzing the inconsistency in these test cases.
To further enhance our top 1 selection process, we introduce Chain-of-Thought (CoT)~\cite{wei_chain--thought_2023} reasoning as an additional validation mechanism. This approach allows the LLM to reason through the inconsistencies and predict the output reference signal for test cases where inconsistencies are found.

\begin{figure}[tb]
    \centering
    \includegraphics[width=0.95\linewidth]{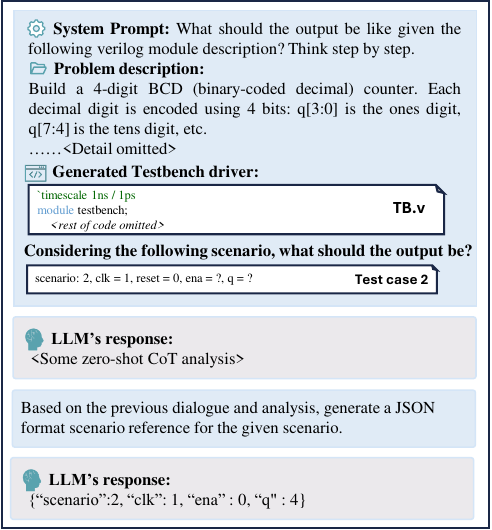}
    \caption{generating reference signals of given scenario from LLM}
    \label{fig:prompt_2}
    \vspace{-0.6cm}
\end{figure}

As illustrated in Fig. \ref{fig:overall_framework}(b), we compare the outputs of two candidate clusters (e.g., cluster 1 and cluster 2, with cluster 1 having a higher score). A test case \(T_{d}\) in which cluster 1 and cluster 2 generate different outputs is identified. This test case, the testbench code we previously generated and the module’s specification, are fed to LLM. Using zero-shot CoT reasoning,  the LLM is prompted to predict the reference output for this specific test case. After this reasoning attempt, we prompt LLM again to summarize its reasoning and return the reference output in JSON format (Ref. in the figure). Fig. \ref{fig:prompt_2} shows an example of this prompt structure. The above attempt of CoT reasoning and summarizing is repeated \(x\) times, generating a set of reference outputs.

If a specified proportion (over $th\%$ percent) of the LLM-reasoned reference signals is identical and aligns with cluster 2's output, we swap cluster 1 with cluster 2, making cluster 2 our top choice. We can carry this same reasoning process on any number of clusters, by comparing cluster 1 and cluster 2, then the new cluster 1 and cluster 3, etc. This process ensures that our final decision of top 1 is not solely based on clustering but reinforced by logical reasoning aligned with the expected behavior of the Verilog module.

\section{Experimental Setup}
\label{sec:ExperimentalSetup}

\subsubsection{Verilog generation benchmark}
The effectiveness of our proposed framework is carried on VerilogEval-Human~\cite{liu_verilogeval_2023}. The dataset contains 156 hand-crafted Verilog generation problems. All simulations of LLM-generated tests and ground truth testbenches (only for validating our top pick) are carried out on iverilog~\cite{williams_steveicarusiverilog_2024}.

\subsubsection{Evaluation Metrics}
We use the original pass@k definition in ~\cite{chen_evaluating_2021} as a basic evaluation method for our system. 
The metric describes the rate of any of the first $k$ code candidates that are functional and pass the human-designed testbench. The baseline of choosing based on random pick is given by:

\begin{equation}
\text{pass@k} := \mathbb{E}_{\text{Problems}} \left[ 1 - \frac{\binom{n-c}{k}}{\binom{n}{k}} \right]
\end{equation}

Where \(n\) stands for the total sample number, \(c\) for the number of codes that passes in $n$ samples. When running the experiment, we have an $n$ of 50 that is on par with our algorithm, to form an accurate baseline estimation fair for comparison. Regarding test cases number $m$, we prompt the language model to give at least 10 test cases. To alleviate the randomness in test case generation, the experiment below is repeated 5 times.


To demonstrate VRank can be applied to different LLMs, we choose LLMs of different sizes, general and domain-specific, open-source and close-source models. 

Our code generation of open-sourced models is carried on 4 NVIDIA A100 GPUs, using Huggingface Transformers~\cite{huggingface_huggingface_nodate} for CodeV-QWen~\cite{zhao_codev_2024} and vLLM framework for Llama-3-70B~\cite{dubey_llama_2024} using 4-bit quantization with AutoAWQ~\cite{lin_awq_2024}. Regarding close-source models, we use OpenAI's latest model, gpt-4o-2024-08-06 and gpt-4o-mini-2024-07-18. All code candidates are generated on the model's default temperature. All simulations are carried on a server with 2 Xeon Gold 6126 CPU and 280 GB RAM.

\section{Experimental Results}
\label{sec:ExperimentalResults}

In this section, we evaluate the performance of our proposed VRank framework using the VerilogEval-Human benchmark. We test its effectiveness across various large language models (LLMs), including two closed-source models (GPT-4o and GPT-4o-mini) and two open-source models (LLama and CodeV). Our experiments are designed to answer the following research questions (RQs):
    \textbf{RQ1:} Can self-consistency of simulation output serve as a reliable indicator for selecting the correct Verilog code?
    \textbf{RQ2:} Does the VRank system improve the accuracy of different models?
    \textbf{RQ3:} How does the performance of our system vary with different sample sizes?
    \textbf{RQ4:} Is there any other loss function we can use to minimize inconsistency?
    \textbf{RQ5:} Can Chain-of-Thought decision provide high-quality inconsistency resolution?

\subsection{RQ1: Validating simulation output consistency}
\label{sec:rq1_cluster_agreement}
The first experiment examined the core assumption of the VRank framework: \textbf{correct solutions exhibit more consistency} among generated code candidates, while incorrect solutions are more diverse. We analyzed 50 Verilog codes generated by GPT-4o across 156 problems, a total of 50*156=7800 samples. If a code sample's simulation output has an exact match on all test cases with any other code samples, we call it has consistency.

As shown in Table \ref{table:agreement_insight}, 56.7\% (4426) of the solutions were correct, and had consistency, whereas only 14.6\% (1135) were incorrect but had consistency. In contrast, candidates without any consistency were more likely to be incorrect, with 28.6\% (2233) incorrect and fewer than 0.1\% (6) correct.

\begin{table}[tb]
\centering
\caption{Candidate number for consistency}
\label{table:agreement_insight}
\begin{tabular}{l|c c}
\hline
\multicolumn{1}{c|}{\textbf{}} & \textbf{has consistency} & \textbf{no consistency} \\ \hline
\textbf{Correct}   & 4426  &6 \\ 
\textbf{Incorrect} & 1135  &2233 \\ \hline
\end{tabular}
\vspace{-0.3cm}
\end{table}

\begin{table}[tb]
\centering
\caption{Top clusters for GPT-4o}
\label{table:example_of_top_picks}
\begin{tabular}{l|c c c c}
\hline
\multicolumn{1}{c|}{\textbf{}} & \textbf{1st} & \textbf{2nd} & \textbf{3rd} & \textbf{left pick} \\ \hline
\textbf{Correct}    & 107   & 6(8) & 0(1) & 0 \\ 
\textbf{Incorrect}  & 6     & 25& 20& N/A \\ \hline
\end{tabular}
\vspace{-0.3cm}
\end{table}

As shown in Fig.~\ref{table:example_of_top_picks}, 113 tasks resulted in at least one cluster containing a correct solution following our clustering and ranking algorithm. Among these, 107 tasks had a correct code candidate ranked in the top 1 cluster, and the remaining 6 had correct candidates ranked second. In parenthesis, we also demonstrate the number of tasks for which the second or third candidate contains a correct code candidate. If the top 2 clusters failed, there was no leftover pick in the cluster we did not choose. This result demonstrated that our method is highly effective, meaning that an engineer would only need to evaluate the top two clusters when using VRank with GPT-4o.

\begin{table*}[tb]
\centering
\begin{threeparttable}
\caption{Comparison of the proposed framework with direct generation baseline}
\label{table:main_result}
\begin{tabular}{l c |c c c|c c c c |c}
\toprule[1.2pt]
\multirow{2}{*}{\textbf{Model}} & \multirow{2}{*}{\textbf{Dataset}} & \multicolumn{3}{c|}{\textbf{Baseline}} & \multicolumn{4}{c|}{\textbf{Our Method}} & \multirow{2}{*}{\textbf{Increase}} \\ \cmidrule{3-9}
 & & \textbf{Pass@1} & \textbf{Pass@2} & \textbf{Pass@3} & \textbf{Pass@1} & \textbf{Pass@2} & \textbf{Pass@3} & \textbf{CoT Reference} &  \\ 
\midrule[1.2pt]
GPT-4o-mini & \multirow{4}{*}{Human} & 48.8\% & 53.2\% & 55.4\% & \textbf{57.5\%} & 58.7\% & 59.9\% & \textbf{58.1\%} & \textbf{9.3\%} \\ 
GPT-4o     &                        & 57.4\% & 62.0\% & 64.4\% & \textbf{66.6\%} & 70.2\% & 70.8\% & \textbf{67.8\%} & \textbf{10.4\%} \\ 
Llama      &                        & 41.8\% & 46.8\% & 49.3\% & \textbf{48.5\%} & 51.9\% & 54.1\% & N/A & \textbf{6.7\%} \\ 
CodeV-Qwen &                        & 32.4\% & 43.7\% & 53.2\% & \textbf{48.1\%} & 53.4\% & 54.6\% & N/A & \textbf{15.7\%} \\ 
\midrule[1.2pt]
GPT-4o-mini &\multirow{4}{*}{CMB(81)} & 68.0\% & 72.5\% & 74.5\% & \textbf{78.0\%} & 78.2\% & 78.4\% & \textbf{78.0\%} & \textbf{10.0\%} \\ 
GPT-4o     &                     & 71.4\% & 75.5\% & 77.5\% & \textbf{77.7\%} & 82.6\% & 82.6\% & \textbf{78.9\%} & \textbf{7.6\%} \\ 
Llama      &                     & 60.2\% & 64.4\% & 66.3\% & \textbf{67.0\%} & 68.6\% & 69.9\% & N/A & \textbf{6.8\%} \\ 
CodeV-Qwen &                     & 34.1\% & 40.5\% & 44.2\% & \textbf{53.7\%} & 57.5\% & 58.2\% & N/A & \textbf{19.6\%} \\ 
\midrule[1.2pt]
GPT-4o-mini & \multirow{4}{*}{SEQ(75)} & 27.5\% & 32.3\% & 34.7\% & \textbf{34.8\%} & 37.6\% & 39.4\% & \textbf{36.6\%} & \textbf{9.1\%} \\ 
GPT-4o&                           & 41.9\% & 47.0\% & 49.9\% & \textbf{54.3\%} & 56.6\% & 57.7\% & \textbf{55.6\%} & \textbf{13.6\%} \\ 
Llama &                             & 21.4\% & 27.3\% & 30.4\% & \textbf{28.0\%} & 33.4\% & 36.6\% & N/A & \textbf{6.6\%} \\ 
CodeV-Qwen &                        & 30.5\% & 38.7\% & 43.2\% & \textbf{42.0\%} & 49.0\% & 50.6\% & N/A & \textbf{11.5\%} \\ 
\bottomrule[1.2pt]
\end{tabular}
\end{threeparttable}

\begin{tablenotes}
\item[] \scriptsize \hfill CoT Reference results are not reported for Llama and CodeV-Qwen since they do not support formalized JSON output.
\end{tablenotes}
\vspace{-0.3cm}
\end{table*}

\subsection{RQ2: Accuracy Improvement Across Models}
\label{sec:rq2_accuracy_improvement}

The performance of VRank across various LLMs is presented in Table \ref{table:main_result}. The baseline comparison was formed by random sampling, reflecting the default method for code generation with LLMs. We report the pass@1 to pass@3 scores for both the baseline and our method, without applying Chain-of-Thought (CoT) reasoning. The pass@1 results after applying CoT reasoning will be discussed in RQ5.

Our experimental results demonstrate that VRank consistently improves accuracy across all tested models, including both closed-source and open-source models. Notably, the pass@1 metric significantly improves when using VRank compared to the baseline.
The most substantial gains in performance were observed with the CodeV-Qwen model. VRank significantly enhanced its capability in generating both combinational (CMB) and sequential (SEQ) circuits, showing a 19.6\% and 11.5\% improvement, respectively. Under our prompting strategy, which may differ from the prompting it sees when finetuning on Verilog, CodeV-Qwen can still produce valid solutions, but experiences much variability in baseline. VRank greatly helps to alleviate the variance and shows the highest improvement among the models tested.

The highest functional correctness was observed with GPT-4o, with the second largest increase. The experimental data shows that VRank enhances the accuracy of code generation, with especially pronounced gains in models with lower initial performance, but remains very effective even for high-performing models like GPT-4o and GPT-4o-mini.

\subsection{RQ3: Impact of Sample Size on Performance}
\label{sec:rq3_sample_size}

\begin{figure*}[tb]
    \centering
    \includegraphics[width=1\linewidth]{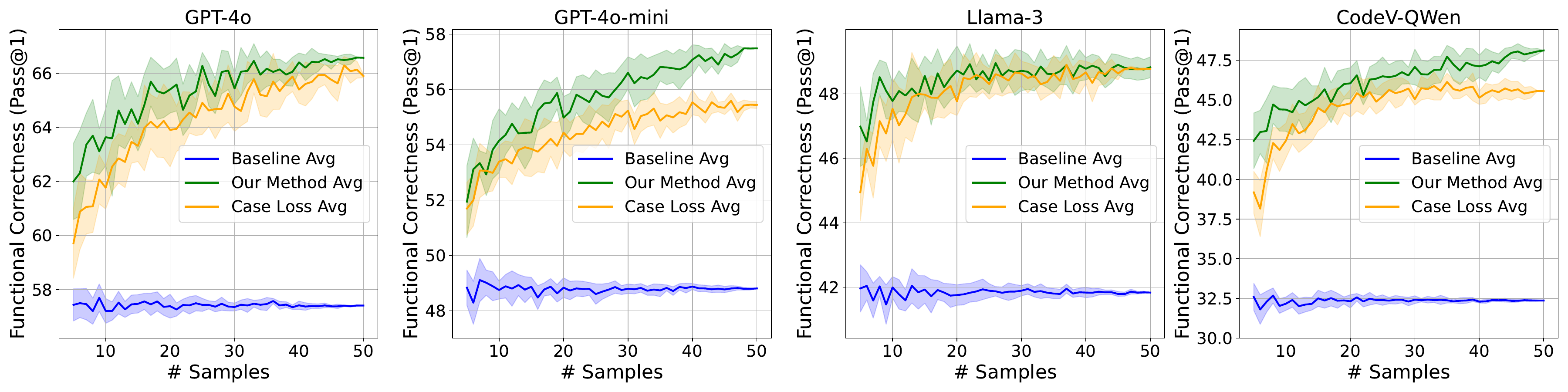}
    \caption{Functional correctness increase as \# Samples increase}
    \label{fig:sample_num}
    \vspace{-0.6cm}
\end{figure*}
To investigate the impact of sample size on VRank’s performance, we evaluated the Pass@1 accuracy of VRank on different LLMs as the number of code candidates varied from 5 to 50. The experiment was repeated 10 times, and we report the average (lines) and the standard deviation (shaded area) of pass@1 in Fig. \ref{fig:sample_num}. VRank consistently (our method) outperformed the baseline, even with as few as five samples. When the sample size reached 20, we observed that approximately 90\% of the performance gain (relative to the 50-sample results) was already achieved. This suggests that engineers could use a smaller sample size when working with more expensive models, balancing computational cost with performance.

\subsection{RQ4: Impact of other Loss design on Performance}
\label{sec:rq4_gpt4_reflection}

To explore the design of the consistency metric, we also explored evaluating consistency score in a case-wise consistency function:
\begin{equation}
R_{case}(c)=n-\sum_{c'\in C}\ell_{case}(c,c')
\end{equation}
with loss 
\begin{equation}
\ell_{case}(c,c')=\frac{1}{m} \sum_{t\in T}\mathbf{1}[c(t)\neq c'(t)]
\end{equation}
The case-wise consistency score measures each code candidate by consistency on each test case. Instead of strict consistency that allows no difference in output, the case-wise score measures the consistency in a 0-1 range. The result is presented in Fig. \ref{fig:sample_num}, labeled Case Loss in the graph. We observe a \(0-2\%\) lower accuracy in the pass@1 ratio with different models and different sample sizes, indicating that a case-wise score function is less favorable in identifying the best code candidate.

\begin{figure}[tb]
    \centering
    \includegraphics[width=0.8\linewidth]{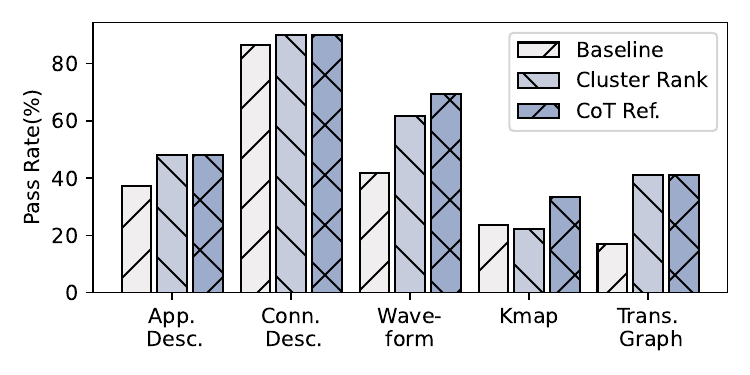}
    \caption{Detail taxonomy pass rate of GPT-4o}
    \label{fig:taxonomy}
    \vspace{-0.6cm}
\end{figure}

\subsection{RQ5: Quality of Inconsistency Resolution with Chain-of-Thought Reasoning}
\label{sec:rq4_gpt4_reflection}
In this final experiment, we evaluate how Chain-of-Thought (CoT) reasoning further enhances the pass@1 accuracy of our framework. We applied CoT reasoning to GPT-4o and GPT-4o-mini on identifying top \(2\) clusters. Both demonstrated the ability to correct previous mistakes by resolving inconsistencies between top-ranked code candidates.
To better understand the CoT reasoning’s impact, we generated a taxonomy of the VerilogEval-Human benchmark inputs, as shown in Fig. \ref{fig:taxonomy}. The inputs were categorized into five types: application description, connection description, waveform description, K-map description, and transition graph. We observed that on more complex reasoning tasks (e.g., application and connection descriptions), CoT reasoning effectively filtered out unstable predictions (\(th\%=80\%, x=5\)), maintaining high accuracy at stage 1 of the process. On more straightforward tasks (e.g., waveform and K-map descriptions), CoT reasoning further improved pass@1 accuracy.

\section{Conclusion}
\label{sec:conclusion}
In this paper, we presented VRank, an automatic framework for improving the pass@1 rate of Verilog code generation using large language models (LLMs). Our approach integrates execution-based clustering, ranking, and Chain-of-Thought (CoT) reasoning to systematically select the most reliable code candidates without relying on human-written testbenches. By using self-consistency on both simulation output and reference signal reasoning, VRank ensures that the selected code is both highly agreed upon and logically consistent with the expected hardware behavior.
The experimental results demonstrate that VRank improves Pass@1 accuracy by an average of 10.5\% across different models. Furthermore, VRank consistently outperforms the baseline, even with a sample size as small as 5. Our framework is efficient and scalable for various applications, providing a powerful tool for enhancing Verilog code generation with LLMs.



\printbibliography

@article{chen_machine_2023,
	title = {Machine Learning in Advanced {IC} Design: A Methodological Survey},
	volume = {40},
	issn = {2168-2364},
	url = {https://ieeexplore.ieee.org/document/9927393},
	doi = {10.1109/MDAT.2022.3216799},
	shorttitle = {Machine Learning in Advanced {IC} Design},
	pages = {17--33},
	number = {1},
	journaltitle = {{IEEE} Design \& Test},
	author = {Chen, Tinghuan and Zhang, Grace Li and Yu, Bei and Li, Bing and Schlichtmann, Ulf},
	urldate = {2025-01-22},
	date = {2023-02},
	note = {Conference Name: {IEEE} Design \& Test},
}

@misc{xu_automated_2024,
	title = {Automated C/C++ Program Repair for High-Level Synthesis via Large Language Models},
	url = {http://arxiv.org/abs/2407.03889},
	doi = {10.48550/arXiv.2407.03889},
	number = {{arXiv}:2407.03889},
	publisher = {{arXiv}},
	author = {Xu, Kangwei and Zhang, Grace Li and Yin, Xunzhao and Zhuo, Cheng and Schlichtmann, Ulf and Li, Bing},
	urldate = {2025-01-22},
	date = {2024-07-04},
	eprinttype = {arxiv},
	eprint = {2407.03889 [eess]},
}

@misc{xu_llm-aided_2024,
	title = {{LLM}-Aided Efficient Hardware Design Automation},
	url = {http://arxiv.org/abs/2410.18582},
	doi = {10.48550/arXiv.2410.18582},
	number = {{arXiv}:2410.18582},
	publisher = {{arXiv}},
	author = {Xu, Kangwei and Qiu, Ruidi and Zhao, Zhuorui and Zhang, Grace Li and Schlichtmann, Ulf and Li, Bing},
	urldate = {2025-01-22},
	date = {2024-10-24},
	eprinttype = {arxiv},
	eprint = {2410.18582 [eess]},
}

@misc{qiu_correctbench_2024,
	title = {{CorrectBench}: Automatic Testbench Generation with Functional Self-Correction using {LLMs} for {HDL} Design},
	url = {http://arxiv.org/abs/2411.08510},
	doi = {10.48550/arXiv.2411.08510},
	series = {{DATE}'25},
	shorttitle = {{CorrectBench}},
	number = {{arXiv}:2411.08510},
	publisher = {{arXiv}},
	author = {Qiu, Ruidi and Zhang, Grace Li and Drechsler, Rolf and Schlichtmann, Ulf and Li, Bing},
	urldate = {2024-12-11},
	date = {2024-11-13},
	eprinttype = {arxiv},
	eprint = {2411.08510 [cs]},
}

@online{williams_steveicarusiverilog_2024,
	title = {steveicarus/iverilog},
	rights = {{GPL}-2.0},
	url = {https://github.com/steveicarus/iverilog},
	author = {Williams, Stephen},
	urldate = {2024-09-18},
	date = {2024-09-17},
	note = {original-date: 2008-05-12T16:57:52Z},
}

@online{huggingface_huggingface_nodate,
	title = {Huggingface Transformers},
	url = {https://huggingface.co/docs/transformers/index},
	author = {Huggingface},
	urldate = {2024-09-22},
}

@misc{wei_chain--thought_2023,
	title = {Chain-of-Thought Prompting Elicits Reasoning in Large Language Models},
	url = {http://arxiv.org/abs/2201.11903},
	doi = {10.48550/arXiv.2201.11903},
	number = {{arXiv}:2201.11903},
	publisher = {{arXiv}},
	author = {Wei, Jason and Wang, Xuezhi and Schuurmans, Dale and Bosma, Maarten and Ichter, Brian and Xia, Fei and Chi, Ed and Le, Quoc and Zhou, Denny},
	urldate = {2024-09-22},
	date = {2023-01-10},
	eprinttype = {arxiv},
	eprint = {2201.11903 [cs]},
}

@inproceedings{lewis_retrieval-augmented_2020,
	title = {Retrieval-Augmented Generation for Knowledge-Intensive {NLP} Tasks},
	volume = {33},
	url = {https://proceedings.neurips.cc/paper/2020/hash/6b493230205f780e1bc26945df7481e5-Abstract.html},
	pages = {9459--9474},
	booktitle = {Advances in Neural Information Processing Systems},
	publisher = {Curran Associates, Inc.},
	author = {Lewis, Patrick and Perez, Ethan and Piktus, Aleksandra and Petroni, Fabio and Karpukhin, Vladimir and Goyal, Naman and Küttler, Heinrich and Lewis, Mike and Yih, Wen-tau and Rocktäschel, Tim and Riedel, Sebastian and Kiela, Douwe},
	urldate = {2024-09-22},
	date = {2020},
}

@inproceedings{thakur_benchmarking_2023,
	title = {Benchmarking Large Language Models for Automated Verilog {RTL} Code Generation},
	url = {https://ieeexplore.ieee.org/abstract/document/10137086},
	doi = {10.23919/DATE56975.2023.10137086},
	eventtitle = {2023 Design, Automation \& Test in Europe Conference \& Exhibition ({DATE})},
	pages = {1--6},
	booktitle = {2023 Design, Automation \& Test in Europe Conference \& Exhibition ({DATE})},
	author = {Thakur, Shailja and Ahmad, Baleegh and Fan, Zhenxing and Pearce, Hammond and Tan, Benjamin and Karri, Ramesh and Dolan-Gavitt, Brendan and Garg, Siddharth},
	urldate = {2024-09-22},
	date = {2023-04},
	note = {{ISSN}: 1558-1101},
}

@inproceedings{he_large_2024,
	location = {New York, {NY}, {USA}},
	title = {Large Language Models for {EDA}: Future or Mirage?},
	isbn = {9798400704178},
	url = {https://doi.org/10.1145/3626184.3639700},
	doi = {10.1145/3626184.3639700},
	series = {{ISPD} '24},
	shorttitle = {Large Language Models for {EDA}},
	pages = {65--66},
	booktitle = {Proceedings of the 2024 International Symposium on Physical Design},
	publisher = {Association for Computing Machinery},
	author = {He, Zhuolun and Yu, Bei},
	urldate = {2024-09-22},
	date = {2024-03-12},
}

@online{gao_autovcoder_2024,
	title = {{AutoVCoder}: A Systematic Framework for Automated Verilog Code Generation using {LLMs}},
	shorttitle = {{AutoVCoder}},
	titleaddon = {{arXiv}.org},
	author = {Gao, Mingzhe and Zhao, Jieru and Lin, Zhe and Ding, Wenchao and Hou, Xiaofeng and Feng, Yu and Li, Chao and Guo, Minyi},
	urldate = {2024-09-12},
	date = {2024-07-21},
	langid = {english},
}

@article{lin_awq_2024,
	title = {{AWQ}: Activation-aware Weight Quantization for On-Device {LLM} Compression and Acceleration},
	volume = {6},
	url = {https://proceedings.mlsys.org/paper_files/paper/2024/hash/42a452cbafa9dd64e9ba4aa95cc1ef21-Abstract-Conference.html},
	shorttitle = {{AWQ}},
	pages = {87--100},
	journaltitle = {Proceedings of Machine Learning and Systems},
	author = {Lin, Ji and Tang, Jiaming and Tang, Haotian and Yang, Shang and Chen, Wei-Ming and Wang, Wei-Chen and Xiao, Guangxuan and Dang, Xingyu and Gan, Chuang and Han, Song},
	urldate = {2024-09-22},
	date = {2024-05-29},
	langid = {english},
}

@inproceedings{qiu_autobench_2024,
	location = {New York, {NY}, {USA}},
	title = {{AutoBench}: Automatic Testbench Generation and Evaluation Using {LLMs} for {HDL} Design},
	isbn = {9798400706998},
	url = {https://dl.acm.org/doi/10.1145/3670474.3685956},
	doi = {10.1145/3670474.3685956},
	series = {{MLCAD} '24},
	shorttitle = {{AutoBench}},
	pages = {1--10},
	booktitle = {Proceedings of the 2024 {ACM}/{IEEE} International Symposium on Machine Learning for {CAD}},
	publisher = {Association for Computing Machinery},
	author = {Qiu, Ruidi and Zhang, Grace Li and Drechsler, Rolf and Schlichtmann, Ulf and Li, Bing},
	urldate = {2024-09-22},
	date = {2024-09-09},
}

@inproceedings{pei_betterv_2024,
	title = {{BetterV}: Controlled Verilog Generation with Discriminative Guidance},
	url = {https://proceedings.mlr.press/v235/pei24e.html},
	shorttitle = {{BetterV}},
	eventtitle = {International Conference on Machine Learning},
	pages = {40145--40153},
	booktitle = {Proceedings of the 41st International Conference on Machine Learning},
	publisher = {{PMLR}},
	author = {Pei, Zehua and Zhen, Huiling and Yuan, Mingxuan and Huang, Yu and Yu, Bei},
	urldate = {2024-09-22},
	date = {2024-07-08},
	langid = {english},
	note = {{ISSN}: 2640-3498},
}

@misc{dubey_llama_2024,
	title = {The Llama 3 Herd of Models},
	url = {http://arxiv.org/abs/2407.21783},
	doi = {10.48550/arXiv.2407.21783},
	number = {{arXiv}:2407.21783},
	publisher = {{arXiv}},
	author = {Dubey, Abhimanyu and Jauhri, Abhinav and Pandey, Abhinav and Kadian, Abhishek and Al-Dahle, Ahmad and Letman, Aiesha and Mathur, Akhil and Schelten, Alan and Yang, Amy and Fan, Angela and Goyal, Anirudh and Hartshorn, Anthony and Yang, Aobo and Mitra, Archi and Sravankumar, Archie and Korenev, Artem and Hinsvark, Arthur and Rao, Arun and Zhang, Aston and Rodriguez, Aurelien and Gregerson, Austen and Spataru, Ava and Roziere, Baptiste and Biron, Bethany and Tang, Binh and Chern, Bobbie and Caucheteux, Charlotte and Nayak, Chaya and Bi, Chloe and Marra, Chris and {McConnell}, Chris and Keller, Christian and Touret, Christophe and Wu, Chunyang and Wong, Corinne and Ferrer, Cristian Canton and Nikolaidis, Cyrus and Allonsius, Damien and Song, Daniel and Pintz, Danielle and Livshits, Danny and Esiobu, David and Choudhary, Dhruv and Mahajan, Dhruv and Garcia-Olano, Diego and Perino, Diego and Hupkes, Dieuwke and Lakomkin, Egor and {AlBadawy}, Ehab and Lobanova, Elina and Dinan, Emily and Smith, Eric Michael and Radenovic, Filip and Zhang, Frank and Synnaeve, Gabriel and Lee, Gabrielle and Anderson, Georgia Lewis and Nail, Graeme and Mialon, Gregoire and Pang, Guan and Cucurell, Guillem and Nguyen, Hailey and Korevaar, Hannah and Xu, Hu and Touvron, Hugo and Zarov, Iliyan and Ibarra, Imanol Arrieta and Kloumann, Isabel and Misra, Ishan and Evtimov, Ivan and Copet, Jade and Lee, Jaewon and Geffert, Jan and Vranes, Jana and Park, Jason and Mahadeokar, Jay and Shah, Jeet and van der Linde, Jelmer and Billock, Jennifer and Hong, Jenny and Lee, Jenya and Fu, Jeremy and Chi, Jianfeng and Huang, Jianyu and Liu, Jiawen and Wang, Jie and Yu, Jiecao and Bitton, Joanna and Spisak, Joe and Park, Jongsoo and Rocca, Joseph and Johnstun, Joshua and Saxe, Joshua and Jia, Junteng and Alwala, Kalyan Vasuden and Upasani, Kartikeya and Plawiak, Kate and Li, Ke and Heafield, Kenneth and Stone, Kevin and El-Arini, Khalid and Iyer, Krithika and Malik, Kshitiz and Chiu, Kuenley and Bhalla, Kunal and Rantala-Yeary, Lauren and van der Maaten, Laurens and Chen, Lawrence and Tan, Liang and Jenkins, Liz and Martin, Louis and Madaan, Lovish and Malo, Lubo and Blecher, Lukas and Landzaat, Lukas and de Oliveira, Luke and Muzzi, Madeline and Pasupuleti, Mahesh and Singh, Mannat and Paluri, Manohar and Kardas, Marcin and Oldham, Mathew and Rita, Mathieu and Pavlova, Maya and Kambadur, Melanie and Lewis, Mike and Si, Min and Singh, Mitesh Kumar and Hassan, Mona and Goyal, Naman and Torabi, Narjes and Bashlykov, Nikolay and Bogoychev, Nikolay and Chatterji, Niladri and Duchenne, Olivier and Çelebi, Onur and Alrassy, Patrick and Zhang, Pengchuan and Li, Pengwei and Vasic, Petar and Weng, Peter and Bhargava, Prajjwal and Dubal, Pratik and Krishnan, Praveen and Koura, Punit Singh and Xu, Puxin and He, Qing and Dong, Qingxiao and Srinivasan, Ragavan and Ganapathy, Raj and Calderer, Ramon and Cabral, Ricardo Silveira and Stojnic, Robert and Raileanu, Roberta and Girdhar, Rohit and Patel, Rohit and Sauvestre, Romain and Polidoro, Ronnie and Sumbaly, Roshan and Taylor, Ross and Silva, Ruan and Hou, Rui and Wang, Rui and Hosseini, Saghar and Chennabasappa, Sahana and Singh, Sanjay and Bell, Sean and Kim, Seohyun Sonia and Edunov, Sergey and Nie, Shaoliang and Narang, Sharan and Raparthy, Sharath and Shen, Sheng and Wan, Shengye and Bhosale, Shruti and Zhang, Shun and Vandenhende, Simon and Batra, Soumya and Whitman, Spencer and Sootla, Sten and Collot, Stephane and Gururangan, Suchin and Borodinsky, Sydney and Herman, Tamar and Fowler, Tara and Sheasha, Tarek and Georgiou, Thomas and Scialom, Thomas and Speckbacher, Tobias and Mihaylov, Todor and Xiao, Tong and Karn, Ujjwal and Goswami, Vedanuj and Gupta, Vibhor and Ramanathan, Vignesh and Kerkez, Viktor and Gonguet, Vincent and Do, Virginie and Vogeti, Vish and Petrovic, Vladan and Chu, Weiwei and Xiong, Wenhan and Fu, Wenyin and Meers, Whitney and Martinet, Xavier and Wang, Xiaodong and Tan, Xiaoqing Ellen and Xie, Xinfeng and Jia, Xuchao and Wang, Xuewei and Goldschlag, Yaelle and Gaur, Yashesh and Babaei, Yasmine and Wen, Yi and Song, Yiwen and Zhang, Yuchen and Li, Yue and Mao, Yuning and Coudert, Zacharie Delpierre and Yan, Zheng and Chen, Zhengxing and Papakipos, Zoe and Singh, Aaditya and Grattafiori, Aaron and Jain, Abha and Kelsey, Adam and Shajnfeld, Adam and Gangidi, Adithya and Victoria, Adolfo and Goldstand, Ahuva and Menon, Ajay and Sharma, Ajay and Boesenberg, Alex and Vaughan, Alex and Baevski, Alexei and Feinstein, Allie and Kallet, Amanda and Sangani, Amit and Yunus, Anam and Lupu, Andrei and Alvarado, Andres and Caples, Andrew and Gu, Andrew and Ho, Andrew and Poulton, Andrew and Ryan, Andrew and Ramchandani, Ankit and Franco, Annie and Saraf, Aparajita and Chowdhury, Arkabandhu and Gabriel, Ashley and Bharambe, Ashwin and Eisenman, Assaf and Yazdan, Azadeh and James, Beau and Maurer, Ben and Leonhardi, Benjamin and Huang, Bernie and Loyd, Beth and De Paola, Beto and Paranjape, Bhargavi and Liu, Bing and Wu, Bo and Ni, Boyu and Hancock, Braden and Wasti, Bram and Spence, Brandon and Stojkovic, Brani and Gamido, Brian and Montalvo, Britt and Parker, Carl and Burton, Carly and Mejia, Catalina and Wang, Changhan and Kim, Changkyu and Zhou, Chao and Hu, Chester and Chu, Ching-Hsiang and Cai, Chris and Tindal, Chris and Feichtenhofer, Christoph and Civin, Damon and Beaty, Dana and Kreymer, Daniel and Li, Daniel and Wyatt, Danny and Adkins, David and Xu, David and Testuggine, Davide and David, Delia and Parikh, Devi and Liskovich, Diana and Foss, Didem and Wang, Dingkang and Le, Duc and Holland, Dustin and Dowling, Edward and Jamil, Eissa and Montgomery, Elaine and Presani, Eleonora and Hahn, Emily and Wood, Emily and Brinkman, Erik and Arcaute, Esteban and Dunbar, Evan and Smothers, Evan and Sun, Fei and Kreuk, Felix and Tian, Feng and Ozgenel, Firat and Caggioni, Francesco and Guzmán, Francisco and Kanayet, Frank and Seide, Frank and Florez, Gabriela Medina and Schwarz, Gabriella and Badeer, Gada and Swee, Georgia and Halpern, Gil and Thattai, Govind and Herman, Grant and Sizov, Grigory and Guangyi and Zhang and Lakshminarayanan, Guna and Shojanazeri, Hamid and Zou, Han and Wang, Hannah and Zha, Hanwen and Habeeb, Haroun and Rudolph, Harrison and Suk, Helen and Aspegren, Henry and Goldman, Hunter and Damlaj, Ibrahim and Molybog, Igor and Tufanov, Igor and Veliche, Irina-Elena and Gat, Itai and Weissman, Jake and Geboski, James and Kohli, James and Asher, Japhet and Gaya, Jean-Baptiste and Marcus, Jeff and Tang, Jeff and Chan, Jennifer and Zhen, Jenny and Reizenstein, Jeremy and Teboul, Jeremy and Zhong, Jessica and Jin, Jian and Yang, Jingyi and Cummings, Joe and Carvill, Jon and Shepard, Jon and {McPhie}, Jonathan and Torres, Jonathan and Ginsburg, Josh and Wang, Junjie and Wu, Kai and U, Kam Hou and Saxena, Karan and Prasad, Karthik and Khandelwal, Kartikay and Zand, Katayoun and Matosich, Kathy and Veeraraghavan, Kaushik and Michelena, Kelly and Li, Keqian and Huang, Kun and Chawla, Kunal and Lakhotia, Kushal and Huang, Kyle and Chen, Lailin and Garg, Lakshya and A, Lavender and Silva, Leandro and Bell, Lee and Zhang, Lei and Guo, Liangpeng and Yu, Licheng and Moshkovich, Liron and Wehrstedt, Luca and Khabsa, Madian and Avalani, Manav and Bhatt, Manish and Tsimpoukelli, Maria and Mankus, Martynas and Hasson, Matan and Lennie, Matthew and Reso, Matthias and Groshev, Maxim and Naumov, Maxim and Lathi, Maya and Keneally, Meghan and Seltzer, Michael L. and Valko, Michal and Restrepo, Michelle and Patel, Mihir and Vyatskov, Mik and Samvelyan, Mikayel and Clark, Mike and Macey, Mike and Wang, Mike and Hermoso, Miquel Jubert and Metanat, Mo and Rastegari, Mohammad and Bansal, Munish and Santhanam, Nandhini and Parks, Natascha and White, Natasha and Bawa, Navyata and Singhal, Nayan and Egebo, Nick and Usunier, Nicolas and Laptev, Nikolay Pavlovich and Dong, Ning and Zhang, Ning and Cheng, Norman and Chernoguz, Oleg and Hart, Olivia and Salpekar, Omkar and Kalinli, Ozlem and Kent, Parkin and Parekh, Parth and Saab, Paul and Balaji, Pavan and Rittner, Pedro and Bontrager, Philip and Roux, Pierre and Dollar, Piotr and Zvyagina, Polina and Ratanchandani, Prashant and Yuvraj, Pritish and Liang, Qian and Alao, Rachad and Rodriguez, Rachel and Ayub, Rafi and Murthy, Raghotham and Nayani, Raghu and Mitra, Rahul and Li, Raymond and Hogan, Rebekkah and Battey, Robin and Wang, Rocky and Maheswari, Rohan and Howes, Russ and Rinott, Ruty and Bondu, Sai Jayesh and Datta, Samyak and Chugh, Sara and Hunt, Sara and Dhillon, Sargun and Sidorov, Sasha and Pan, Satadru and Verma, Saurabh and Yamamoto, Seiji and Ramaswamy, Sharadh and Lindsay, Shaun and Lindsay, Shaun and Feng, Sheng and Lin, Shenghao and Zha, Shengxin Cindy and Shankar, Shiva and Zhang, Shuqiang and Zhang, Shuqiang and Wang, Sinong and Agarwal, Sneha and Sajuyigbe, Soji and Chintala, Soumith and Max, Stephanie and Chen, Stephen and Kehoe, Steve and Satterfield, Steve and Govindaprasad, Sudarshan and Gupta, Sumit and Cho, Sungmin and Virk, Sunny and Subramanian, Suraj and Choudhury, Sy and Goldman, Sydney and Remez, Tal and Glaser, Tamar and Best, Tamara and Kohler, Thilo and Robinson, Thomas and Li, Tianhe and Zhang, Tianjun and Matthews, Tim and Chou, Timothy and Shaked, Tzook and Vontimitta, Varun and Ajayi, Victoria and Montanez, Victoria and Mohan, Vijai and Kumar, Vinay Satish and Mangla, Vishal and Albiero, Vítor and Ionescu, Vlad and Poenaru, Vlad and Mihailescu, Vlad Tiberiu and Ivanov, Vladimir and Li, Wei and Wang, Wenchen and Jiang, Wenwen and Bouaziz, Wes and Constable, Will and Tang, Xiaocheng and Wang, Xiaofang and Wu, Xiaojian and Wang, Xiaolan and Xia, Xide and Wu, Xilun and Gao, Xinbo and Chen, Yanjun and Hu, Ye and Jia, Ye and Qi, Ye and Li, Yenda and Zhang, Yilin and Zhang, Ying and Adi, Yossi and Nam, Youngjin and Yu and Wang and Hao, Yuchen and Qian, Yundi and He, Yuzi and Rait, Zach and {DeVito}, Zachary and Rosnbrick, Zef and Wen, Zhaoduo and Yang, Zhenyu and Zhao, Zhiwei},
	urldate = {2024-09-19},
	date = {2024-08-15},
	eprinttype = {arxiv},
	eprint = {2407.21783 [cs]},
}

@inproceedings{papineni_bleu_2002,
	location = {Philadelphia, Pennsylvania, {USA}},
	title = {Bleu: a Method for Automatic Evaluation of Machine Translation},
	url = {https://aclanthology.org/P02-1040},
	doi = {10.3115/1073083.1073135},
	pages = {311--318},
	booktitle = {Proceedings of the 40th Annual Meeting of the Association for Computational Linguistics},
	publisher = {Association for Computational Linguistics},
	author = {Papineni, Kishore and Roukos, Salim and Ward, Todd and Zhu, Wei-Jing},
	editor = {Isabelle, Pierre and Charniak, Eugene and Lin, Dekang},
	date = {2002-07},
}

@misc{chang_chipgpt_2023,
	title = {{ChipGPT}: How far are we from natural language hardware design},
	url = {http://arxiv.org/abs/2305.14019},
	doi = {10.48550/arXiv.2305.14019},
	shorttitle = {{ChipGPT}},
	number = {{arXiv}:2305.14019},
	publisher = {{arXiv}},
	author = {Chang, Kaiyan and Wang, Ying and Ren, Haimeng and Wang, Mengdi and Liang, Shengwen and Han, Yinhe and Li, Huawei and Li, Xiaowei},
	urldate = {2024-09-16},
	date = {2023-06-19},
	eprinttype = {arxiv},
	eprint = {2305.14019 [cs]},
}

@inproceedings{pearce_dave_2020,
	location = {New York, {NY}, {USA}},
	title = {{DAVE}: Deriving Automatically Verilog from English},
	isbn = {978-1-4503-7519-1},
	url = {https://dl.acm.org/doi/10.1145/3380446.3430634},
	doi = {10.1145/3380446.3430634},
	series = {{MLCAD} '20},
	shorttitle = {{DAVE}},
	pages = {27--32},
	booktitle = {Proceedings of the 2020 {ACM}/{IEEE} Workshop on Machine Learning for {CAD}},
	publisher = {Association for Computing Machinery},
	author = {Pearce, Hammond and Tan, Benjamin and Karri, Ramesh},
	urldate = {2024-09-15},
	date = {2020-11-16},
}

@article{li_competition-level_2022,
	title = {Competition-Level Code Generation with {AlphaCode}},
	volume = {378},
	issn = {0036-8075, 1095-9203},
	url = {http://arxiv.org/abs/2203.07814},
	doi = {10.1126/science.abq1158},
	pages = {1092--1097},
	number = {6624},
	journaltitle = {Science},
	shortjournal = {Science},
	author = {Li, Yujia and Choi, David and Chung, Junyoung and Kushman, Nate and Schrittwieser, Julian and Leblond, Rémi and Eccles, Tom and Keeling, James and Gimeno, Felix and Lago, Agustin Dal and Hubert, Thomas and Choy, Peter and d'Autume, Cyprien de Masson and Babuschkin, Igor and Chen, Xinyun and Huang, Po-Sen and Welbl, Johannes and Gowal, Sven and Cherepanov, Alexey and Molloy, James and Mankowitz, Daniel J. and Robson, Esme Sutherland and Kohli, Pushmeet and de Freitas, Nando and Kavukcuoglu, Koray and Vinyals, Oriol},
	urldate = {2024-09-12},
	date = {2022-12-09},
	eprinttype = {arxiv},
	eprint = {2203.07814 [cs]},
}

@inproceedings{fiscus_post-processing_1997,
	title = {A post-processing system to yield reduced word error rates: Recognizer output voting error reduction ({ROVER})},
	pages = {347--354},
	booktitle = {1997 {IEEE} Workshop on Automatic Speech Recognition and Understanding Proceedings},
	publisher = {{IEEE}},
	author = {Fiscus, Jonathan G},
	date = {1997},
}

@inproceedings{goel_segmental_2000,
	title = {Segmental minimum Bayes-risk {ASR} voting strategies},
	booktitle = {Sixth International Conference on Spoken Language Processing},
	publisher = {Citeseer},
	author = {Goel, Vaibhava and Kumar, Shankar and Byrne, William},
	date = {2000},
}

@inproceedings{kumar_minimum_2004,
	title = {Minimum bayes-risk decoding for statistical machine translation},
	pages = {169--176},
	booktitle = {Proceedings of the Human Language Technology Conference of the North American Chapter of the Association for Computational Linguistics: {HLT}-{NAACL} 2004},
	author = {Kumar, Shankar and Byrne, Bill},
	date = {2004},
}

@misc{chen_evaluating_2021,
	title = {Evaluating Large Language Models Trained on Code},
	url = {http://arxiv.org/abs/2107.03374},
	doi = {10.48550/arXiv.2107.03374},
	number = {{arXiv}:2107.03374},
	publisher = {{arXiv}},
	author = {Chen, Mark and Tworek, Jerry and Jun, Heewoo and Yuan, Qiming and Pinto, Henrique Ponde de Oliveira and Kaplan, Jared and Edwards, Harri and Burda, Yuri and Joseph, Nicholas and Brockman, Greg and Ray, Alex and Puri, Raul and Krueger, Gretchen and Petrov, Michael and Khlaaf, Heidy and Sastry, Girish and Mishkin, Pamela and Chan, Brooke and Gray, Scott and Ryder, Nick and Pavlov, Mikhail and Power, Alethea and Kaiser, Lukasz and Bavarian, Mohammad and Winter, Clemens and Tillet, Philippe and Such, Felipe Petroski and Cummings, Dave and Plappert, Matthias and Chantzis, Fotios and Barnes, Elizabeth and Herbert-Voss, Ariel and Guss, William Hebgen and Nichol, Alex and Paino, Alex and Tezak, Nikolas and Tang, Jie and Babuschkin, Igor and Balaji, Suchir and Jain, Shantanu and Saunders, William and Hesse, Christopher and Carr, Andrew N. and Leike, Jan and Achiam, Josh and Misra, Vedant and Morikawa, Evan and Radford, Alec and Knight, Matthew and Brundage, Miles and Murati, Mira and Mayer, Katie and Welinder, Peter and {McGrew}, Bob and Amodei, Dario and {McCandlish}, Sam and Sutskever, Ilya and Zaremba, Wojciech},
	urldate = {2024-09-04},
	date = {2021-07-14},
	eprinttype = {arxiv},
	eprint = {2107.03374 [cs]},
}

@misc{liu_chipnemo_2024,
	title = {{ChipNeMo}: Domain-Adapted {LLMs} for Chip Design},
	url = {http://arxiv.org/abs/2311.00176},
	doi = {10.48550/arXiv.2311.00176},
	shorttitle = {{ChipNeMo}},
	number = {{arXiv}:2311.00176},
	publisher = {{arXiv}},
	author = {Liu, Mingjie and Ene, Teodor-Dumitru and Kirby, Robert and Cheng, Chris and Pinckney, Nathaniel and Liang, Rongjian and Alben, Jonah and Anand, Himyanshu and Banerjee, Sanmitra and Bayraktaroglu, Ismet and Bhaskaran, Bonita and Catanzaro, Bryan and Chaudhuri, Arjun and Clay, Sharon and Dally, Bill and Dang, Laura and Deshpande, Parikshit and Dhodhi, Siddhanth and Halepete, Sameer and Hill, Eric and Hu, Jiashang and Jain, Sumit and Jindal, Ankit and Khailany, Brucek and Kokai, George and Kunal, Kishor and Li, Xiaowei and Lind, Charley and Liu, Hao and Oberman, Stuart and Omar, Sujeet and Pasandi, Ghasem and Pratty, Sreedhar and Raiman, Jonathan and Sarkar, Ambar and Shao, Zhengjiang and Sun, Hanfei and Suthar, Pratik P. and Tej, Varun and Turner, Walker and Xu, Kaizhe and Ren, Haoxing},
	urldate = {2024-09-04},
	date = {2024-04-04},
	eprinttype = {arxiv},
	eprint = {2311.00176 [cs]},
}

@inproceedings{liu_verilogeval_2023,
	title = {{VerilogEval}: Evaluating Large Language Models for Verilog Code Generation},
	booktitle = {2023 {IEEE}/{ACM} International Conference on Computer-Aided Design ({ICCAD})},
	author = {Liu, Mingjie and Pinckney, Nathaniel and Khailany, Brucek and Ren, Haoxing},
	date = {2023},
}

@misc{liu_rtlcoder_2024,
	title = {{RTLCoder}: Outperforming {GPT}-3.5 in Design {RTL} Generation with Our Open-Source Dataset and Lightweight Solution},
	url = {http://arxiv.org/abs/2312.08617},
	doi = {10.48550/arXiv.2312.08617},
	shorttitle = {{RTLCoder}},
	number = {{arXiv}:2312.08617},
	publisher = {{arXiv}},
	author = {Liu, Shang and Fang, Wenji and Lu, Yao and Zhang, Qijun and Zhang, Hongce and Xie, Zhiyao},
	urldate = {2024-09-04},
	date = {2024-02-20},
	eprinttype = {arxiv},
	eprint = {2312.08617 [cs]},
}

@misc{openai_gpt-4_2024,
	title = {{GPT}-4 Technical Report},
	url = {http://arxiv.org/abs/2303.08774},
	doi = {10.48550/arXiv.2303.08774},
	number = {{arXiv}:2303.08774},
	publisher = {{arXiv}},
	author = {{OpenAI} and Achiam, Josh and Adler, Steven and Agarwal, Sandhini and Ahmad, Lama and Akkaya, Ilge and Aleman, Florencia Leoni and Almeida, Diogo and Altenschmidt, Janko and Altman, Sam and Anadkat, Shyamal and Avila, Red and Babuschkin, Igor and Balaji, Suchir and Balcom, Valerie and Baltescu, Paul and Bao, Haiming and Bavarian, Mohammad and Belgum, Jeff and Bello, Irwan and Berdine, Jake and Bernadett-Shapiro, Gabriel and Berner, Christopher and Bogdonoff, Lenny and Boiko, Oleg and Boyd, Madelaine and Brakman, Anna-Luisa and Brockman, Greg and Brooks, Tim and Brundage, Miles and Button, Kevin and Cai, Trevor and Campbell, Rosie and Cann, Andrew and Carey, Brittany and Carlson, Chelsea and Carmichael, Rory and Chan, Brooke and Chang, Che and Chantzis, Fotis and Chen, Derek and Chen, Sully and Chen, Ruby and Chen, Jason and Chen, Mark and Chess, Ben and Cho, Chester and Chu, Casey and Chung, Hyung Won and Cummings, Dave and Currier, Jeremiah and Dai, Yunxing and Decareaux, Cory and Degry, Thomas and Deutsch, Noah and Deville, Damien and Dhar, Arka and Dohan, David and Dowling, Steve and Dunning, Sheila and Ecoffet, Adrien and Eleti, Atty and Eloundou, Tyna and Farhi, David and Fedus, Liam and Felix, Niko and Fishman, Simón Posada and Forte, Juston and Fulford, Isabella and Gao, Leo and Georges, Elie and Gibson, Christian and Goel, Vik and Gogineni, Tarun and Goh, Gabriel and Gontijo-Lopes, Rapha and Gordon, Jonathan and Grafstein, Morgan and Gray, Scott and Greene, Ryan and Gross, Joshua and Gu, Shixiang Shane and Guo, Yufei and Hallacy, Chris and Han, Jesse and Harris, Jeff and He, Yuchen and Heaton, Mike and Heidecke, Johannes and Hesse, Chris and Hickey, Alan and Hickey, Wade and Hoeschele, Peter and Houghton, Brandon and Hsu, Kenny and Hu, Shengli and Hu, Xin and Huizinga, Joost and Jain, Shantanu and Jain, Shawn and Jang, Joanne and Jiang, Angela and Jiang, Roger and Jin, Haozhun and Jin, Denny and Jomoto, Shino and Jonn, Billie and Jun, Heewoo and Kaftan, Tomer and Kaiser, Łukasz and Kamali, Ali and Kanitscheider, Ingmar and Keskar, Nitish Shirish and Khan, Tabarak and Kilpatrick, Logan and Kim, Jong Wook and Kim, Christina and Kim, Yongjik and Kirchner, Jan Hendrik and Kiros, Jamie and Knight, Matt and Kokotajlo, Daniel and Kondraciuk, Łukasz and Kondrich, Andrew and Konstantinidis, Aris and Kosic, Kyle and Krueger, Gretchen and Kuo, Vishal and Lampe, Michael and Lan, Ikai and Lee, Teddy and Leike, Jan and Leung, Jade and Levy, Daniel and Li, Chak Ming and Lim, Rachel and Lin, Molly and Lin, Stephanie and Litwin, Mateusz and Lopez, Theresa and Lowe, Ryan and Lue, Patricia and Makanju, Anna and Malfacini, Kim and Manning, Sam and Markov, Todor and Markovski, Yaniv and Martin, Bianca and Mayer, Katie and Mayne, Andrew and {McGrew}, Bob and {McKinney}, Scott Mayer and {McLeavey}, Christine and {McMillan}, Paul and {McNeil}, Jake and Medina, David and Mehta, Aalok and Menick, Jacob and Metz, Luke and Mishchenko, Andrey and Mishkin, Pamela and Monaco, Vinnie and Morikawa, Evan and Mossing, Daniel and Mu, Tong and Murati, Mira and Murk, Oleg and Mély, David and Nair, Ashvin and Nakano, Reiichiro and Nayak, Rajeev and Neelakantan, Arvind and Ngo, Richard and Noh, Hyeonwoo and Ouyang, Long and O'Keefe, Cullen and Pachocki, Jakub and Paino, Alex and Palermo, Joe and Pantuliano, Ashley and Parascandolo, Giambattista and Parish, Joel and Parparita, Emy and Passos, Alex and Pavlov, Mikhail and Peng, Andrew and Perelman, Adam and Peres, Filipe de Avila Belbute and Petrov, Michael and Pinto, Henrique Ponde de Oliveira and Michael and Pokorny and Pokrass, Michelle and Pong, Vitchyr H. and Powell, Tolly and Power, Alethea and Power, Boris and Proehl, Elizabeth and Puri, Raul and Radford, Alec and Rae, Jack and Ramesh, Aditya and Raymond, Cameron and Real, Francis and Rimbach, Kendra and Ross, Carl and Rotsted, Bob and Roussez, Henri and Ryder, Nick and Saltarelli, Mario and Sanders, Ted and Santurkar, Shibani and Sastry, Girish and Schmidt, Heather and Schnurr, David and Schulman, John and Selsam, Daniel and Sheppard, Kyla and Sherbakov, Toki and Shieh, Jessica and Shoker, Sarah and Shyam, Pranav and Sidor, Szymon and Sigler, Eric and Simens, Maddie and Sitkin, Jordan and Slama, Katarina and Sohl, Ian and Sokolowsky, Benjamin and Song, Yang and Staudacher, Natalie and Such, Felipe Petroski and Summers, Natalie and Sutskever, Ilya and Tang, Jie and Tezak, Nikolas and Thompson, Madeleine B. and Tillet, Phil and Tootoonchian, Amin and Tseng, Elizabeth and Tuggle, Preston and Turley, Nick and Tworek, Jerry and Uribe, Juan Felipe Cerón and Vallone, Andrea and Vijayvergiya, Arun and Voss, Chelsea and Wainwright, Carroll and Wang, Justin Jay and Wang, Alvin and Wang, Ben and Ward, Jonathan and Wei, Jason and Weinmann, C. J. and Welihinda, Akila and Welinder, Peter and Weng, Jiayi and Weng, Lilian and Wiethoff, Matt and Willner, Dave and Winter, Clemens and Wolrich, Samuel and Wong, Hannah and Workman, Lauren and Wu, Sherwin and Wu, Jeff and Wu, Michael and Xiao, Kai and Xu, Tao and Yoo, Sarah and Yu, Kevin and Yuan, Qiming and Zaremba, Wojciech and Zellers, Rowan and Zhang, Chong and Zhang, Marvin and Zhao, Shengjia and Zheng, Tianhao and Zhuang, Juntang and Zhuk, William and Zoph, Barret},
	urldate = {2024-09-03},
	date = {2024-03-04},
	eprinttype = {arxiv},
	eprint = {2303.08774 [cs]},
}

@misc{ho_verilogcoder_2024,
	title = {{VerilogCoder}: Autonomous Verilog Coding Agents with Graph-based Planning and Abstract Syntax Tree ({AST})-based Waveform Tracing Tool},
	url = {http://arxiv.org/abs/2408.08927},
	doi = {10.48550/arXiv.2408.08927},
	shorttitle = {{VerilogCoder}},
	number = {{arXiv}:2408.08927},
	publisher = {{arXiv}},
	author = {Ho, Chia-Tung and Ren, Haoxing and Khailany, Brucek},
	urldate = {2024-08-22},
	date = {2024-08-15},
	eprinttype = {arxiv},
	eprint = {2408.08927 [cs]},
	note = {version: 1},
}

@misc{cui_origenenhancing_2024,
	title = {{OriGen}:Enhancing {RTL} Code Generation with Code-to-Code Augmentation and Self-Reflection},
	url = {http://arxiv.org/abs/2407.16237},
	doi = {10.48550/arXiv.2407.16237},
	shorttitle = {{OriGen}},
	number = {{arXiv}:2407.16237},
	publisher = {{arXiv}},
	author = {Cui, Fan and Yin, Chenyang and Zhou, Kexing and Xiao, Youwei and Sun, Guangyu and Xu, Qiang and Guo, Qipeng and Song, Demin and Lin, Dahua and Zhang, Xingcheng and Yun and Liang},
	urldate = {2024-08-09},
	date = {2024-07-23},
	eprinttype = {arxiv},
	eprint = {2407.16237 [cs]},
}

@misc{zhao_codev_2024,
	title = {{CodeV}: Empowering {LLMs} for Verilog Generation through Multi-Level Summarization},
	url = {http://arxiv.org/abs/2407.10424},
	doi = {10.48550/arXiv.2407.10424},
	shorttitle = {{CodeV}},
	number = {{arXiv}:2407.10424},
	publisher = {{arXiv}},
	author = {Zhao, Yang and Huang, Di and Li, Chongxiao and Jin, Pengwei and Nan, Ziyuan and Ma, Tianyun and Qi, Lei and Pan, Yansong and Zhang, Zhenxing and Zhang, Rui and Zhang, Xishan and Du, Zidong and Guo, Qi and Hu, Xing and Chen, Yunji},
	urldate = {2024-08-02},
	date = {2024-07-20},
	eprinttype = {arxiv},
	eprint = {2407.10424 [cs]},
	note = {version: 4},
}

@inproceedings{wang_self-consistency_2022,
	title = {Self-Consistency Improves Chain of Thought Reasoning in Language Models},
	url = {https://openreview.net/forum?id=1PL1NIMMrw},
	eventtitle = {The Eleventh International Conference on Learning Representations},
	author = {Wang, Xuezhi and Wei, Jason and Schuurmans, Dale and Le, Quoc V. and Chi, Ed H. and Narang, Sharan and Chowdhery, Aakanksha and Zhou, Denny},
	urldate = {2024-07-25},
	date = {2022-09-29},
	langid = {english},
}

@misc{thakur_autochip_2024,
	title = {{AutoChip}: Automating {HDL} Generation Using {LLM} Feedback},
	url = {http://arxiv.org/abs/2311.04887},
	doi = {10.48550/arXiv.2311.04887},
	shorttitle = {{AutoChip}},
	number = {{arXiv}:2311.04887},
	publisher = {{arXiv}},
	author = {Thakur, Shailja and Blocklove, Jason and Pearce, Hammond and Tan, Benjamin and Garg, Siddharth and Karri, Ramesh},
	urldate = {2024-07-16},
	date = {2024-06-04},
	eprinttype = {arxiv},
	eprint = {2311.04887 [cs]},
}

@misc{tsai_rtlfixer_2024,
	title = {{RTLFixer}: Automatically Fixing {RTL} Syntax Errors with Large Language Models},
	url = {http://arxiv.org/abs/2311.16543},
	doi = {10.48550/arXiv.2311.16543},
	shorttitle = {{RTLFixer}},
	number = {{arXiv}:2311.16543},
	publisher = {{arXiv}},
	author = {Tsai, Yun-Da and Liu, Mingjie and Ren, Haoxing},
	urldate = {2024-04-18},
	date = {2024-02-07},
	eprinttype = {arxiv},
	eprint = {2311.16543 [cs]},
}

@inproceedings{blocklove_chip-chat_2023,
	title = {Chip-Chat: Challenges and Opportunities in Conversational Hardware Design},
	url = {http://arxiv.org/abs/2305.13243},
	doi = {10.1109/MLCAD58807.2023.10299874},
	shorttitle = {Chip-Chat},
	pages = {1--6},
	booktitle = {2023 {ACM}/{IEEE} 5th Workshop on Machine Learning for {CAD} ({MLCAD})},
	author = {Blocklove, Jason and Garg, Siddharth and Karri, Ramesh and Pearce, Hammond},
	urldate = {2024-03-11},
	date = {2023-09-10},
	langid = {english},
	eprinttype = {arxiv},
	eprint = {2305.13243 [cs]},
}

@article{chang_improving_nodate,
	title = {Improving Large Language Model Hardware Generating Quality through Post-{LLM} Search},
	author = {Chang, Kaiyan and Ren, Haimeng and Wang, Mengdi and Liang, Shengwen and Han, Yinhe and Li, Huawei and Li, Xiaowei and Wang, Ying},
	langid = {english},
}

\end{document}